\documentclass[conference]{IEEEtran}
\IEEEoverridecommandlockouts
\usepackage{amsmath,amsfonts}
\usepackage[linesnumbered]{algorithm2e}
\usepackage{array}
\usepackage[caption=false,font=normalsize,labelfont=sf,textfont=sf]{subfig}
\usepackage{textcomp}
\usepackage{stfloats}
\usepackage{url,hyperref}
\usepackage{verbatim}
\usepackage{graphicx}
\usepackage{cite}
\usepackage{todonotes}
\usepackage{amssymb}
\usepackage{booktabs}
\usepackage{svg}
\usepackage{multirow,booktabs}
\usepackage{subfig}
\usepackage{tcolorbox}
\usepackage{colortbl}
\usepackage{bm,upgreek}
\usepackage{enumitem}

\newtcolorbox{myquote}{colback=white, colframe=gray, width=\columnwidth, before skip=1em, left=0.5em, right=0.5em, boxrule=0.5mm, arc=0mm, outer arc=0mm, boxsep=0pt}

\usepackage[compact]{titlesec}
\titlespacing{\section}{0pt}{1ex}{0.5ex}
\titlespacing{\subsection}{0pt}{0.5ex}{0ex}
\titlespacing{\subsubsection}{0pt}{0.5ex}{0ex}
\def\BibTeX{{\rm B\kern-.05em{\sc i\kern-.025em b}\kern-.08em
    T\kern-.1667em\lower.7ex\hbox{E}\kern-.125emX}}
\begin{document}

\title{A Data-Driven Method for Microgrid System Identification: Physically Consistent Sparse Identification of Nonlinear Dynamics
\thanks{This work was supported by Natural Sciences and Engineering Research
Council (NSERC) and the Fonds de Recherche du Quebec-Nature et technologies under Grants FRQ-NT PR-298827, FRQ-NT 2023-NOVA-314338, FRQ-NT 344058 and Prix Grands Sages Ashok Vijh en électrochimie.
The authors acknowledge the use of ChatGPT (OpenAI), for grammar correction, language refinement, and stylistic suggestions. All original content, ideas, and technical contributions are the sole work of the authors.}
}

\author{\IEEEauthorblockN{Mohan Du and Xiaozhe Wang}
\IEEEauthorblockA{Department of Electrical and Computer Engineering, 
McGill University, 
Montreal, QC H3A 2K6, Canada \\
mohan.du@mail.mcgill.ca, xiaozhe.wang2@mcgill.ca}

}

\maketitle

\begin{abstract}
Microgrids (MGs) play a crucial role in utilizing distributed energy resources (DERs) like solar and wind power, enhancing the sustainability and flexibility of modern power systems. 
\textcolor{black}{However, the inherent variability in MG topology, power flow, and DER operating modes poses significant challenges to the accurate system identification of MGs, which is crucial for designing robust control strategies and ensuring MG stability.}
This paper proposes a Physically Consistent Sparse Identification of Nonlinear Dynamics (PC-SINDy) method for accurate MG system identification.
By leveraging an analytically derived library of candidate functions, PC-SINDy extracts accurate dynamic models using only phasor measurement unit (PMU) data. Simulations on a 4-bus system demonstrate that PC-SINDy can reliably and accurately predict frequency trajectories under large disturbances, including scenarios not encountered during the identification/training phase, even when using noisy, low-sampled PMU data.
\end{abstract}

\begin{IEEEkeywords}
Microgrid, distributed energy resource (DER), grid forming, grid following, Koopman theory, sparse identification of nonlinear dynamics (SINDy), system identification. 
\end{IEEEkeywords}

\section{Introduction}

Microgrids (MGs) play a critical role in integrating distributed energy resources (DERs) like solar panels, wind turbines, and energy storage into power systems to meet carbon-neutral goals \cite{gong2023d}. By operating in both grid-connected and islanded modes, they enhance renewable energy flexibility \cite{nandakumar2023}. However, MGs face stability challenges due to the stochastic nature of DER outputs and inertia-less converter characteristics \cite{she2023}. Hierarchical control is widely adopted to address these issues by stabilizing converters, correcting long-term deviations, and optimizing power flow through primary, secondary, and tertiary control, respectively \cite{she2023}.

This study focuses on the secondary control that corrects frequency deviation and maintains frequency stability of MGs.  
Existing model-based methods, like small-signal models for centralized \cite{zhang2016} and decentralized \cite{bidram2013} approaches, rely on precise DER and MG models, which are often unavailable due to constant switching and volatile DERs in MGs  \cite{gong2023b}. Model-free methods, such as PI control \cite{lu2014} and consensus-based control \cite{ullah2021}, face challenges with tuning and scalability in heterogeneous MGs \cite{gong2023b}. To address these limitations, data-driven techniques have been proposed to identify system dynamics for secondary control. Data-driven linear identification approaches \cite{madani2021} may not work for large disturbances \cite{gong2023d}. Machine learning (ML) methods, such as 
neural networks (NNs) \cite{kamwa1996}, offer solutions but are constrained by scalability and data requirements.
Reinforcement learning (RL) \cite{du2020} shows promise in control applications \cite{she2023}, yet lacks interpretability and adaptability to new dynamics \cite{gong2023d}.

Koopman theory is gaining attention for transforming nonlinear systems into linear ones using Koopman observables—nonlinear functions defining a new state space \cite{kutz2016m}. 
However, selecting effective observables remains challenging \cite{kutz2016m}. 
The Sparse Identification of Nonlinear Dynamics (SINDy) method exploits the inherent sparsity of many physical systems by constructing a library of candidate functions to efficiently capture system dynamics. This makes SINDy an effective tool for approximating Koopman operators. 
In \cite{cai2023}, SINDy was applied to detect forced oscillations using PMU data. 
\textcolor{black}{\cite{chen2021f} further integrates sparse regression with physics-informed neural networks to identify governing equations directly from measurements.}
In MG applications, the authors of \cite{ma2024} derived the analytical candidate functions for a three-bus MG model but required extensive grid parameters, making it less practical. 
\cite{nandakumar2023} used sinusoidal and polynomial functions to model DER-based MG dynamics; however, this approach may struggle to accurately capture system behavior during unexpected large disturbances.
Inspired by the fact that sinusoidal-driven angle dynamics may emerge when subject to large perturbations in low-inertia power systems, 
\cite{gong2023, gong2023b} employed angle sinusoidal 
functions in their candidate library and used observer Kalman filter identification to iteratively refine system dynamics, improving control adaptability. Despite advancements in handling nonlinearity and large disturbances, this iterative process is computationally intensive. 

In this paper, we propose a Physically Consistent SINDy (PC-SINDy) method for identifying models of MG systems. The PC-SINDy constructs an analytically derived library of candidate functions, enabling it to estimate the true physical model using only PMU data.
Unlike iterative methods, the identified model directly corresponds to the true system dynamics, allowing accurate prediction of frequency trajectories during large disturbances without iterative refining. 
The main advantages of the proposed PC-SINDy are as follows: 
\begin{enumerate}[label=\arabic*)]
    \item 
    The proposed PC-SINDy method 
    requires no prior knowledge about MG network or DER parameters. 
    \item The proposed PC-SINDy, requiring no online refining, is computationally efficient for online applications. 
    \item Simulation results on a 4-bus system show that PC-SINDy can accurately predict the frequency trajectories 
    using practical PMU data (IEEE Std C37.118.1-2011 \cite{IEEE2011}), even for large disturbances not encountered during the identification/training process.  
\end{enumerate}

\section{Modeling of Frequency Dynamics of DER-Based MG}\label{sec:DER-dynamics}

In this section, we present the modeling of frequency dynamics for grid-forming (GFM) and grid-following (GFL) converters within MGs. 
DER dynamics in MGs can be described as a set of ordinary differential equations (ODEs) \cite{cai2023}: 
\begin{equation}\label{equ:dynamic-system}
    \dot{\boldsymbol{x}}=\boldsymbol{f}(\boldsymbol{x}, \boldsymbol{u}),
\end{equation} 
where \(\boldsymbol{x} =\begin{bmatrix}x_1  & x_2  & \dots & x_N \end{bmatrix}^{\top}\) is an \(N\)-dimensional state vector; \(\boldsymbol{u} = \begin{bmatrix}u_1 & u_2 & \dots & u_{N_{\boldsymbol{u}}} \end{bmatrix}^{\top}\) is an \(N_{\boldsymbol{u}}\)-dimensional control vector; 
and \(\boldsymbol{f}=\begin{bmatrix}f_1  & f_2  & \dots &f_N \end{bmatrix}^{\top}\) is a vector of governing equations for the system \cite{kutz2016d}.

\begin{figure}[htbp!]
    \centering
    \includegraphics[width=0.9\linewidth]{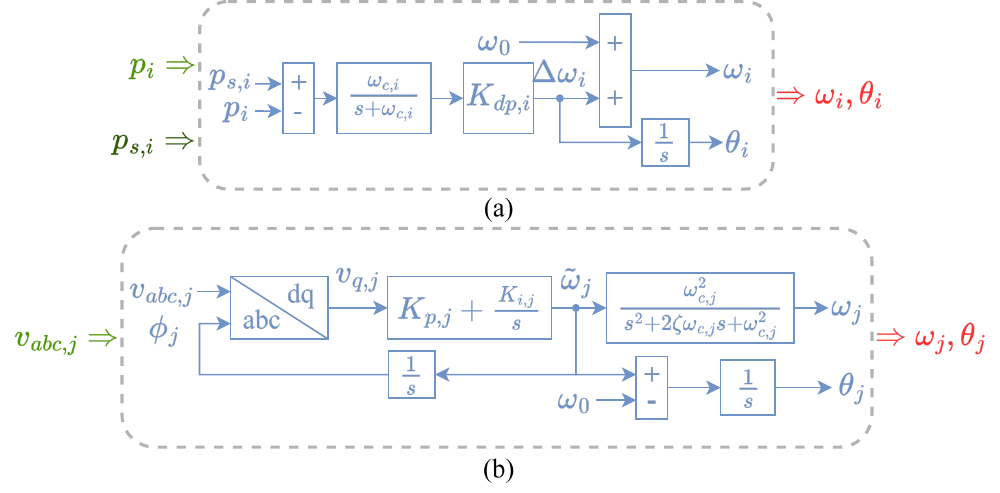}
    \caption{(a) Droop control of GFM converters and (b) three-phase PLL diagrams of GFL converters.}
    \label{fig:circuit-and-control}
\end{figure}

\subsection{GFM Converter}

Fig. \ref{fig:circuit-and-control}(a) presents a commonly used droop control with a low-pass filter in a GFM converter \(i\). 
Droop control in GFM \(i\) uses a first-order low-pass filter \(H_{i}(s) = \frac{\omega_{c,i}}{s + \omega_{c,i}}\) to suppress high-frequency noise, where \(\omega_{c,i}\) is the cut-off frequency. It takes the measured active power \(p_{i}\) and setpoint \(p_{s,i}\) as inputs to generate the reference angular frequency \(\omega_{i}\) and phase angle \(\theta_{i}\): 
\begin{equation}\label{equ:gfm-droop}
    \begin{aligned}
        & \dot{\theta}_{i}=\omega_{i}- \omega_{0}, \\
        & \dot{\omega}_{i} = - \omega_{c,i} \omega_{i} - \omega_{c,i} K_{dp,i} p_{i} + \omega_{c,i} \omega_{0} + \omega_{c,i} K_{dp,i}p_{s,i}, 
    \end{aligned}
\end{equation}
where 
\textcolor{black}{\(\theta_i\) and \(\omega_i\) are state variables; the rated angular frequency \(\omega_0\), \(p_i\), and \(p_{s,i}\) \textcolor{black}{can be termed as control inputs};}
and \(K_{dp,i}\) is the slope of the frequency-active power droop curve. 

\subsection{GFL Converter}
GFL converters use phase-locked loops (PLLs) for synchronization with system frequency and angle \cite{Yazdani2010}. Accurate PLL modeling is crucial for analyzing GFL dynamics, with the three-phase PLL being commonly used, as shown in Fig. \ref{fig:circuit-and-control}(b). 

The three-phase PLL in a GFL converter \(j\) takes the measured three-phase voltage \(v_{abc,j}\) as input and outputs the estimated angular frequency \(\omega_{j}\) and phase angle \(\theta_{j}\). This voltage is first transformed into the q-axis component \(v_{q,j}\) using a dq transformation:
\begin{equation}\label{equ:3p-pll-vq}\footnotesize 
    v_{q,j} = \frac{2}{3} \left( v_{a,j} \sin(\phi_{j}) + v_{b,j} \sin\left(\phi_{j} - \frac{2\pi}{3}\right) + v_{c,j} \sin\left(\phi_{j} + \frac{2\pi}{3}\right) \right), 
\end{equation}
where \(\phi_{j}=\omega_{0}t+\theta_{j}\) is the instantaneous phase. 
Afterward, \(v_{q,j}\) is fed into a PI controller to estimate the angular frequency \(\tilde{\omega}_{j}\):
\begin{equation}\label{equ:3p-pll-omega}
    \tilde{\omega}_{j} = K_{p,j} v_{q,j} + K_{i,j} \int v_{q,j}, 
\end{equation}
where \(K_{p,j}\) and \(K_{i,j}\) are the proportional and integral gains of the PI controller in GFL \(j\), respectively. 
The estimated angular frequency \(\tilde{\omega}_{j}\) determines the phase angle \(\theta_{j}\) following: 
\begin{equation}
    \theta_{j}=\int (\tilde{\omega}_{j}-\omega_{0}). 
\end{equation} 
The angular frequency \(\tilde{\omega}_{j}\) is typically filtered to produce \(\omega_{j}\) \cite{Yazdani2010}. A second-order low-pass filter is commonly applied to mitigate this noise while preserving robustness \cite{matlab-pll3p}:
\begin{equation}\label{equ:3p-pll-filter}
    \frac{1}{\omega_{c,j}^2}\ddot{\omega}_{j} + \frac{2\zeta_{j}}{\omega_{c,j}}\dot{\omega}_{j} + \omega_{j} = \tilde{\omega}_{j}, 
\end{equation}
where \(\omega_{c,j}\) is the natural frequency of the second-order low-pass filter; and \(\zeta_{j}\) is the damping ratio. 

Combining \eqref{equ:3p-pll-vq}-\eqref{equ:3p-pll-filter} yields the PLL dynamics as:
\begin{equation}\footnotesize\label{equ:3p-pll}
    \begin{aligned}
        & \dot{\theta}_{j} = K_{p,j} v_{q,j} + K_{i,j} \int v_{q,j}-\omega_{0}, \\ 
        & \ddot{\omega}_{j} = -2\zeta_{j}\omega_{c,j} \dot{\omega}_{j} - \omega_{c,j}^2 \omega_{j} + \omega_{c,j}^2 K_{p,j} v_{q,j} + \omega_{c,j}^2 K_{i,j} \int v_{q,j},
    \end{aligned}
\end{equation}
\textcolor{black}{where \(\theta_j\), \textcolor{black}{$\omega_j$, \(\dot{\omega}_j\)}, are state variables; and \(\omega_0\) and \(v_{q,j}\) \textcolor{black}{can be termed as} control inputs. }

\section{Sparse Identification of Nonlinear Dynamics}\label{sec:sindy}

In practical applications, many nonlinear dynamical systems with control \eqref{equ:dynamic-system} including DER-based power systems, often exhibit only a few dominant nonlinear terms in their governing equations \cite{brunton2016b}.
The sparsity of governing equations allows approximation by a linear combination of a few candidate functions \cite{cai2023}.
In light of this, \cite{brunton2016b} introduced the SINDy method, which leverages sparsity techniques to identify governing equations directly from measurements, ensuring robust and accurate system identification.

Let \(\boldsymbol{X} \in \mathbb{R}^{M \times N}\) denote the state measurements, with rows corresponding to time points and columns to states, and let \(\boldsymbol{U} \in \mathbb{R}^{M \times N_{\boldsymbol{u}}}\) represent the control inputs over the time interval \(t = [t_1, t_2, \dots, t_M]\). The time derivatives \(\dot{\boldsymbol{X}}\), either directly measured or numerically computed, can be expressed using SINDy theory as a linear combination of candidate functions \cite{brunton2016b}:
\begin{equation} \label{equ:sindy}
    \dot{\boldsymbol{X}}=\Theta(\boldsymbol{X},\boldsymbol{U})\Xi.  
\end{equation}
The library of candidate functions 
\begin{equation}
    \Theta(\boldsymbol{X},\boldsymbol{U})=
    \begin{bmatrix} 
        \theta_1(\boldsymbol{X},\boldsymbol{U}) & \cdots & \theta_P(\boldsymbol{X},\boldsymbol{U}) 
    \end{bmatrix}
\end{equation}
consists of potential functions representing the dominant terms of the governing equations \(\boldsymbol{f}\); 
\(P\) is the total number of functions in the library; 
and the coefficient matrix \(\Xi= \begin{bmatrix} 
        \boldsymbol{\xi}_1 & \boldsymbol{\xi}_2 &\cdots & \boldsymbol{\xi}_N 
    \end{bmatrix}\) consists of sparse vectors \(\boldsymbol{\xi}_i \in \mathbb{R}^{P}\), \(i\in \{1, 2, \dots, N\}\), to be determined. 

Once the coefficient matrix \(\Xi\) is obtained, SINDy could map the original nonlinear system \eqref{equ:dynamic-system} to a low-dimensional linear space spanned by candidate functions
\begin{equation}\label{equ:sindy-dynamic}
    \dot{\boldsymbol{x}}=\left(\Theta(\boldsymbol{x}^{\top}, \boldsymbol{u}^{\top}){\Xi}\right)^{\top}. 
\end{equation}

Accurate linearization of nonlinear systems relies on designing suitable candidate functions \(\Theta(\boldsymbol{X}, \boldsymbol{U})\) \cite{brunton2016b, kutz2016d}. In power systems, these are typically constants, polynomials, and sinusoidal functions \cite{gong2023b, cai2023}. Alternatively, ML methods can implicitly represent candidate functions \cite{gong2023d}. However, both approaches often fail to capture the true governing equations. In Section \ref{sec:methodology}, we derive candidate functions analytically from DER dynamics, enabling SINDy to accurately identify the equations governing DER-based MGs.

\section{The proposed physically consistent SINDy for system identification and frequency regulation} \label{sec:methodology}

This section first demonstrates that the frequency dynamics of DERs in MGs can be expressed in the form of \eqref{equ:sindy-dynamic} (Section \ref{sec:method-candidates}). Building on this, we introduce a physical-consistent SINDy (PC-SINDy) approach for accurate system identification and prediction (Section \ref{sec:method-identify}). The proposed model captures the underlying physical dynamics, allowing precise frequency prediction even under previously unseen disturbances.

\subsection{Analytically Designed Candidate Functions for Capturing DER Dynamics}\label{sec:method-candidates}

The frequency dynamics of the converter GFM \(i\) \eqref{equ:gfm-droop} can be expressed in the SINDy form 
\begin{equation}\label{equ:gfm-droop-sindy}
        \dot{\boldsymbol{X}}_{i}=\Theta_{i}(\boldsymbol{X}_{i}, \boldsymbol{U}_{i}) \Xi_{i}. 
\end{equation}
by defining \(\dot{\boldsymbol{X}}_{i}\), \(\Theta_{i}(\boldsymbol{X}_{i}, \boldsymbol{U}_{i})\), and \(\Xi_{i}\) according to
\begin{equation}\label{equ:sindy-droop}
    \underbrace{\begin{bmatrix} \dot{\theta}_{i} & \dot{\omega}_{i} \end{bmatrix} }_{\dot{\boldsymbol{X}}_{i}}
    =
    \underbrace{\begin{bmatrix} \omega_{i} & \omega_{0} & p_{i} & p_{s,i} \end{bmatrix} }_{\Theta_{i}(\boldsymbol{X}_{i}, \boldsymbol{U}_{i})}
    \underbrace{\begin{bmatrix} 
        1 & -\omega_{c,i} \\ 
        -1 & \omega_{c,i} \\ 
        0 & -\omega_{c,i} K_{dp,i} \\ 
        0 & \omega_{c,i} K_{dp,i} 
    \end{bmatrix}}_{\Xi_{i}}. 
\end{equation}
Equation \eqref{equ:gfm-droop-sindy} effectively maps the droop control dynamics \eqref{equ:gfm-droop} onto a linear space defined by the \(\Theta_{i}(\boldsymbol{X}_{i}, \boldsymbol{U}_{i})\). The values of these functions can be directly obtained or calculated from PMU data (e.g., \(\omega_i\), \(p_i\)) or known setpoints (e.g., \(p_{s,i}\)).

If DER \(i\) is a GFL converter, its PLL dynamics \eqref{equ:3p-pll} can be reformulated in the SINDy form \eqref{equ:gfm-droop-sindy} by defining:
\begin{equation}\small
    \underbrace{\begin{bmatrix} 
        \dot{\theta}_{i} \!&\! \dot{\omega}_{i} \!&\! \ddot{\omega}_{i} \end{bmatrix} }_{\dot{\boldsymbol{X}}_{i}}
    = 
    \underbrace{\begin{bmatrix} 
        \omega_{i} \!&\!  \omega_{0} \!&\!  \dot{\omega}_{i} \!&\!  v_{q,i} \!&\!  \int v_{q,i} 
    \end{bmatrix}} _{\Theta_{i}(\boldsymbol{X}_{i}, \boldsymbol{U}_{i})}
    \underbrace{\begin{bmatrix} 
    0 \!&\! 0 \!&\! - \omega_{c,i}^2 \\ -1 \!&\! 0 \!&\! 0 \\ 0 \!&\! 1 \!&\! -2\zeta_{i}\omega_{c,i} \\ K_{p,i} \!&\! 0 \!&\! \omega_{c,i}^2 K_{p,i} \\ K_{i,i} \!&\! 0 \!&\! \omega_{c,i}^2 K_{i,i} 
    \end{bmatrix}}_{\Xi_{i}}. 
\end{equation}

For a microgrid with \(N_D\) DERs, the combined dynamics of all DERs can be expressed in the SINDy form \eqref{equ:sindy}
where
\begin{subequations}\small\label{equ:sindy-mg}
    \begin{align}
        &\dot{\boldsymbol{X}} = \begin{bmatrix} \dot{\boldsymbol{X}}_{1} & \cdots & \dot{\boldsymbol{X}}_{N_D} \end{bmatrix}, \label{equ:sindy-xdot-all}\\
        &\Theta(\boldsymbol{X}, \boldsymbol{U}) = \begin{bmatrix}
        \Theta_{1}(\boldsymbol{X}_{1}, \boldsymbol{U}_{1}) & \cdots & \Theta_{N_D}(\boldsymbol{X}_{N_D}, \boldsymbol{U}_{N_D}) \end{bmatrix}, \label{equ:sindy-library-all}\\
        &\Xi = \operatorname{diag}(\Xi_{1}, \, \cdots, \, \Xi_{N_D}). \label{equ:sindy-xi-all}
    \end{align}
\end{subequations}
The operator \(\operatorname{diag}\) constructs a block matrix with the specified matrices along its diagonal.

In \eqref{equ:sindy-mg}, \(\dot{\bm{X}}\) and \(\Theta(\bm{X,U})\) can be \textcolor{black}{either directly obtained or numerically computed within PMU before being reported}, while \(\Xi\) contains controller parameters that are often inaccessible to system operators. The next subsection introduces how to extract \(\Xi\) purely from PMU measurements.

\subsection{The PC-SINDy method for identifying the true system}  \label{sec:method-identify}
\begin{figure*}[htbp!]
    \centering
    \includegraphics[width=\linewidth]{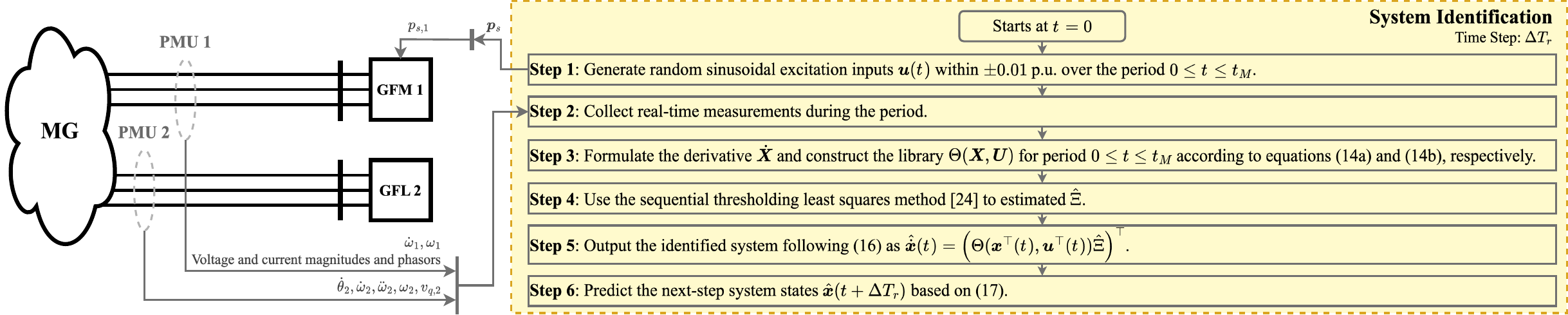}
    \caption{Flowchart of the PC-SINDy-based Identification method for the true physical system.}
    \label{fig:alg-identify}
\end{figure*}

Fig. \ref{fig:alg-identify} illustrates the system identification process using the proposed PC-SINDy method. First, continuous excitation inputs \(\boldsymbol{u}(t)\) are generated as a sum of sinusoidal waves with varying amplitudes and frequencies to stimulate the system. The magnitude of \(\boldsymbol{u}(t)\) is scaled within \(\pm 0.01\) p.u. Each input produces \(M\) data points over the time span \(t_M\). These signals are injected into all GFM converters to excite small system dynamics.

Real-time PMU measurements are then collected in \textbf{Step 2} and are used to construct \(\dot{\boldsymbol{X}}\) and \(\Theta(\boldsymbol{X}, \boldsymbol{U})\) in \textbf{Step 3}. 
In \textbf{Step 4}, the sequential thresholding least squares method \cite{brunton2016b} is used to estimate the coefficient matrix \(\Xi\) from \(\dot{\boldsymbol{X}}\) and \(\Theta(\boldsymbol{X}, \boldsymbol{U})\) by solving the optimization problem
\begin{equation}\label{equ:sindy-objective}
    \hat{\Xi} = \arg \min_{\Xi} \left\| \dot{\boldsymbol{X}} - \Theta(\boldsymbol{X}, \boldsymbol{U}) \Xi \right\|_2.
\end{equation}
Finally, in \textbf{Step 5}, the linearized system model is constructed using the candidate function library from \eqref{equ:sindy-library-all} and the estimated \(\hat{\Xi}\) from \eqref{equ:sindy-objective}: 
\begin{equation}\label{equ:reconstruct-dynamic}
    \hat{\dot{\boldsymbol{x}}}(t)=\left(\Theta(\boldsymbol{x}^{\top}(t), \boldsymbol{u}^{\top}(t))\hat{\Xi}\right)^{\top}.
\end{equation}
Next, in \textbf{Step 6}, the real-time data \(\dot{\boldsymbol{x}}(t)\) and \(\Theta(\boldsymbol{x}^{\top}(t), \boldsymbol{u}^{\top}(t))\) are used to predict the next-step system states \(\dot{\boldsymbol{x}}(t+\Delta T_{r})\) by the Euler's method as in \eqref{equ:sindy-pred}:
\begin{equation}\label{equ:sindy-pred}
    \begin{aligned}
        & \hat{{\boldsymbol{x}}}(t+\Delta T_{r})=\boldsymbol{x}(t)+\hat{\dot{\boldsymbol{x}}}(t) \Delta T_{r}, 
    \end{aligned}
\end{equation}
where \(\Delta T_{r}\) is the time step. 

\textbf{Remarks:} 
\begin{itemize}
    \item \textcolor{black}{The PMU reporting rate in Fig. \ref{fig:alg-identify} is set to \(F_r = 120\) Hz per IEEE standard C37.118.1-2011 \cite{IEEE2011}, corresponding to a time step of \(\Delta T_r \approx 8.33\) ms.}
    \item The excitation process in \textbf{Step 1} runs for \(t_{M} = 10\) s, yielding \(M = 1201\) time points. This has minimal impact on normal operations and requires minimal data.
    \item \textcolor{black}{PMU noise following IEEE C37.118.1-2011 \cite{IEEE2011} and \cite{gong2023b} is considered in \textbf{Step 2} and during the calculation of \(\dot{\boldsymbol{X}}\) and \(\Theta(\boldsymbol{X}, \boldsymbol{U})\).}
    \item The proposed algorithm in \textbf{Step 4} completes in under 1 second using Matlab R2024b on an Intel i7-7700 CPU with 16 GB RAM.
\end{itemize}

\section{Case Study}\label{sec:simulation}

\begin{figure}[htbp]
    \centering
    \includegraphics[width=0.8\linewidth]{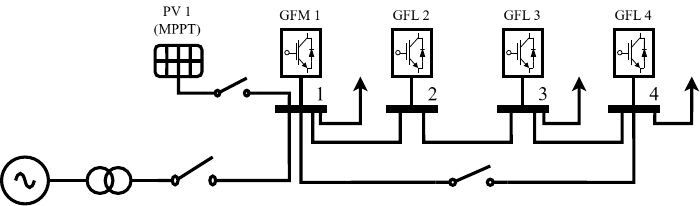}
    \caption{The 4-bus MG.} 
    \label{fig:MG-circuit}
\end{figure}

In this section, the proposed \textbf{PC-SINDy} method is applied to identify DER dynamics and predict system states in MGs. The method’s performance is evaluated under noise and low sampling rates, and is compared with a benchmark approach that uses SINDy with an intuitively developed library of candidate functions \cite{nandakumar2023}.
We consider a 4-bus MG modeled in MATLAB Simulink (see Fig. \ref{fig:MG-circuit}), comprising GFM and GFL converters, a photovoltaic (PV) farm operating in maximum power point tracking mode, and loads with both constant and stochastic power. The converters and their control systems, including inner loops and filters, are modeled to capture electromagnetic transients. The MG can operate in both grid-connected and islanded modes via a breaker and transformer. The rated frequency is set to \(f_0 = 60\) Hz, with remaining parameters as specified in \cite{gong2023b}.

\subsection{System Identification by PC-SINDy Using Noisy Measurements with Low-Sampling Rate}\label{sec:sim-various-measurements}

In \textbf{Step 1}, continuous random inputs \(\boldsymbol{u}(t)\) are generated with a duration of \(t_{M} = 10\) s, ranging within \(\pm 0.01\) p.u., to excite the physical system. 
Next, in \textbf{Step 2}, PMU measurements are collected with noise levels and sampling rates compliant with IEEE C37.118.1-2011 \cite{IEEE2011}, which are then used to construct \(\dot{\boldsymbol{X}}\) and \(\Theta(\boldsymbol{X}, \boldsymbol{U})\) in \textbf{Step 3}. Finally, in \textbf{Step 4}, the sequential thresholding least squares method \cite{brunton2016b} is applied to solve the optimization problem \eqref{equ:sindy-objective} using the constructed data.

\begin{figure}[htbp]
    \centering
    \includegraphics[width=0.8\linewidth]{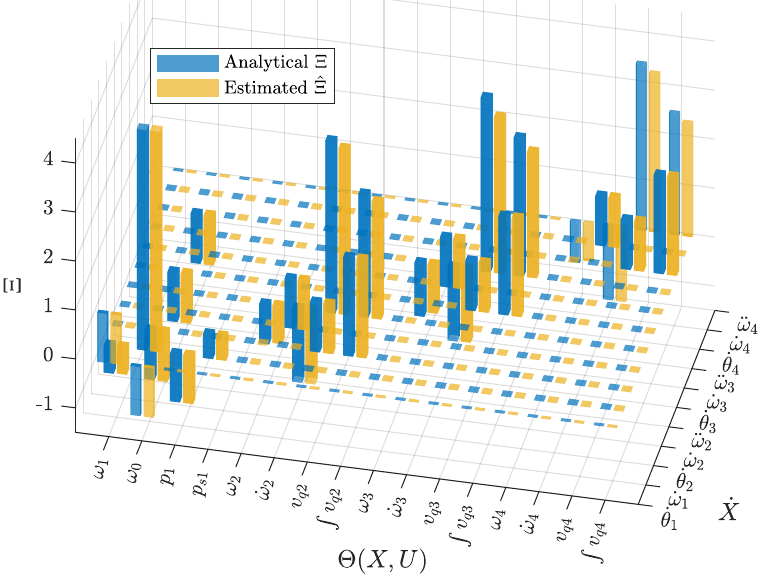}
    \caption{Comparison between analytical \(\Xi\) and SINDy-estimated \(\hat{\Xi}\) using low-quality measurements.}
    \label{fig:Xi-noisy}
\end{figure}

Fig. \ref{fig:Xi-noisy} compares the estimated \(\hat{\Xi}\) from \textbf{Step 4} with the analytical \(\Xi\). While \(\hat{\Xi}\) closely matches \(\Xi\), slight differences arise due to PMU measurement noise and errors in constructing the library. 
The correlation coefficients \(\rho_{\boldsymbol{\xi}_i, \hat{\boldsymbol{\xi}}_i}\), shown in Table \ref{tab:corr_Xi}, are nearly one, indicating that the estimated coefficient vector \(\hat{\boldsymbol{\xi}}_i\) accurately captures the direction of the analytical one \(\boldsymbol{\xi}_i\). The discrepancies are primarily in magnitude \(\frac{\|\hat{\boldsymbol{\xi}}_i\|}{\|{\boldsymbol{\xi}}_i\|}\), due to noise, but these do not significantly impact prediction performance, as shown in Section \ref{sec:sim-prediction}.

\begin{table}[ht]\scriptsize
    \centering
    \caption{Correlation coefficients and ratio between vectors \(\Xi_i\) and \(\hat{\Xi}_i\).}
\begin{tabular}{crcccccccccc}
\toprule
\(\dot{x}_i\) &  \(\dot{\theta}_{1} \)&\( \dot{\omega}_{1}     \)&\( \dot{\theta}_{2} \)&\( \dot{\omega}_{2}     \)&\( \ddot{\omega}_{2}      \)&\( \dot{\theta}_{3} \)&\( \dot{\omega}_{3}     \)&\( \ddot{\omega}_{3}     \)&\( \dot{\theta}_{4} \)&\( \dot{\omega}_{4}     \)&\( \ddot{\omega}_{4}\) \\
\midrule
\(\rho_{\boldsymbol{\xi}_i,\hat{\boldsymbol{\xi}}_i}\) & 1     & .99 & 1     & 1     & 1     & 1     & 1     & 1     & 1     & 1     & 1 \\
\(\frac{\|\hat{\boldsymbol{\xi}}_i\|}{\|{\boldsymbol{\xi}}_i\|}\) & 1     & .99 & .97 & 1     & .95 & .96 & 1     & .97 & .92 & 1     & .94 \\
\bottomrule
\end{tabular}%
    \label{tab:corr_Xi}
\end{table}

\subsection{One-Step-Ahead Frequency Prediction by PC-SINDy}\label{sec:sim-prediction}

In \textbf{Step 5}, the estimated coefficient matrix \(\hat{\Xi}\) from \textbf{Step 4} is used for dynamic prediction. The validation set, spanning \(t \in [10, 13]\) s, includes significant load changes and topology shifts not encountered during system identification. 
At \(t = 10.5\) s, the load at bus 3 increases by 0.7 p.u., followed by an increase in PV 1 generation by 0.4 p.u. at \(t = 11\) s. Subsequently, the AC grid connects at \(t = 11.5\) s and disconnects at \(t = 12\) s, with line 14 linking to the MG at \(t = 12.5\) s.

\begin{figure}[htbp]
    \centering
    \includegraphics[width=\linewidth]{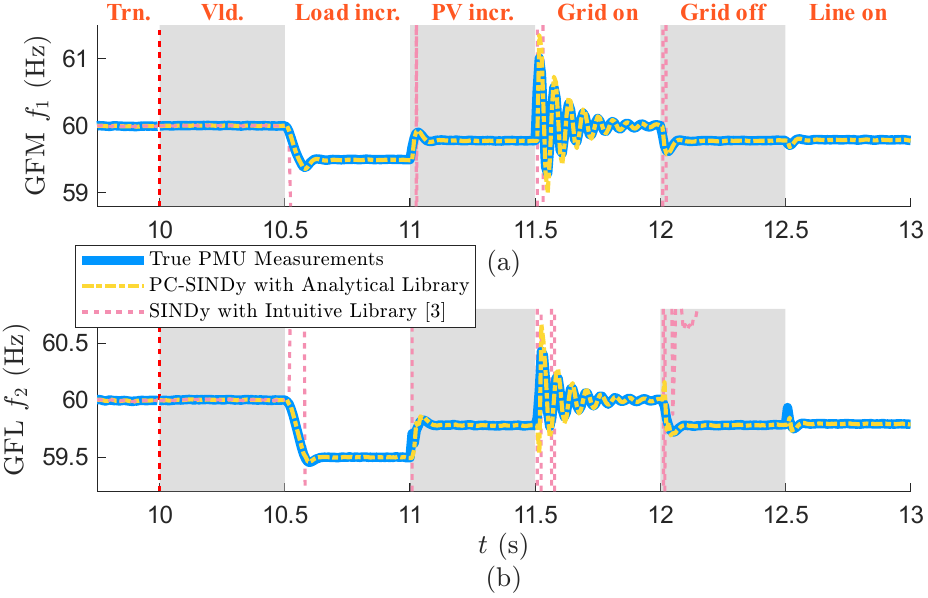}
    \caption{Prediction of the GFM frequency \(f_1\) and the GFL frequency \(f_2\) following \eqref{equ:sindy-pred} with the proposed analytical library and intuitive library \cite{nandakumar2023}. }
    \label{fig:4bus-prediction}
\end{figure}

As a benchmark, we employ the SINDy method with a library of intuitively selected functions, including polynomials and sinusoids of \(\boldsymbol{X}\) and \(\boldsymbol{U}\) \cite{nandakumar2023}.
The corresponding coefficient matrix is extracted from PMU measurements and then used for predictions on the validation set.

Fig.~\ref{fig:4bus-prediction}(a) and (b) display frequency predictions for GFM 1 and GFL 2 using PMU measurements in \textbf{Step 5}. The solid blue lines represent true PMU-measured frequencies \(f_1\) and \(f_2\), while the yellow dash-dotted and pink dashed lines show predictions from the PC-SINDy and intuitive-library-based SINDy \cite{nandakumar2023}, respectively.
Both methods accurately predict frequencies during the quasi-steady state \(t \in [10, 10.5)\) s, where conditions align with training data. However, after the large transient at \(t = 10.5\) s, the intuitive-library-based model diverges beyond the plot range, while PC-SINDy continues to track frequencies accurately, even during untrained transients. For instance, following a 0.7 p.u. load increase at \(t = 10.5\) s, the intuitive model's trajectory deviates significantly, and it fails to track true measurements during topology changes starting at \(t = 11.5\) s.

\section{Conclusion}\label{sec:conclusion}
This paper proposed a data-driven PC-SINDy method for MG system identification. Leveraging an analytically developed candidate function library, PC-SINDy can estimate the true system frequency dynamics without prior information on MG network parameters or DER characteristics. 
Simulations on a 4-bus system showed that PC-SINDy accurately estimates the coefficient matrix for the true system, including ROCOF and second-order frequency derivatives, using only practical PMU data.
PC-SINDy 
can accurately predict the frequency trajectories 
even during large disturbances that are not seen during the training/identification process, 
significantly outperforming the benchmark SINDy method based on intuitive libraries, particularly in noisy, low-resolution conditions.
\textcolor{black}{Future work will extend PC-SINDy to larger MGs and more complex dynamics.}

\bibliography{main.bib}
\bibliographystyle{IEEEtran}

\end{document}